\documentclass{INTERSPEECH2023}

\interspeechcameraready

\usepackage{hyperref}
\usepackage{multirow}
\usepackage{multicol}
\usepackage{comment}
\usepackage{subfig}
\usepackage{amsmath}
\usepackage{amssymb}
\usepackage[vlined]{algorithm2e}
\usepackage{url}
\usepackage{xcolor}
\usepackage{float}
\usepackage{soul}

\title{\textit{Transforming the Embeddings}: A Lightweight Technique for Speech Emotion Recognition Tasks}

\name{Orchid Chetia Phukan$^1$, Arun Balaji Buduru$^1$, Rajesh Sharma$^1$$^,$$^2$}
\address{
  $^1$IIIT-Delhi, India\\
  $^2$University of Tartu, Estonia}
\email{\{orchidp, arunb\}@iiitd.ac.in, rajesh.sharma@ut.ee}

\begin{document}

\maketitle
 
\begin{abstract}
Speech emotion recognition (SER) is a field that has drawn a lot of attention due to its applications in diverse fields. A current trend in methods used for SER is to leverage embeddings from pre-trained models (PTMs) as input features to downstream models. However, the use of embeddings from speaker recognition PTMs hasn't garnered much focus in comparison to other PTM embeddings. To fill this gap and in order to understand the efficacy of speaker recognition PTM embeddings, we perform a comparative analysis of five PTM embeddings. Among all, x-vector embeddings performed the best possibly due to its training for speaker recognition leading to capturing various components of speech such as tone, pitch, etc. Our modeling approach which utilizes x-vector embeddings and mel-frequency cepstral coefficients (MFCC) as input features is the most lightweight approach while achieving comparable accuracy to previous state-of-the-art (SOTA) methods in the CREMA-D benchmark. 

\begin{comment}
Papers for Reference:
1. Speech Emotion Recognition with Multi-task Learning (Interspeech 2021)
2. Speech Emotion Recognition based on Attention Weight Correction
Using Word-level Confidence Measure (Interspeech 2021)
3. Emotion Recognition from Speech Using Wav2vec 2.0 Embeddings (Interspeech 2021)
4. SPEECH EMOTION RECOGNITION USING SELF-SUPERVISED FEATURES (ICASSP 2022)
5. For F1-SCORE for CREMA-D comparison, refer paper (A Comparison Between Convolutional and
Transformer Architectures for Speech Emotion
Recognition)
\end{comment}
\end{abstract}

\section{Introduction}

Human beings express different emotions under different circumstances. %This can be an outburst from a calamitous circumstance or a joyful reaction to a triumph. 
Emotions serve as routes of communication between humans. Emotional connection helps humans to communicate more effectively and as social beings, helps humans understand each other in a better way, celebrate joyful moments together, and be a supporting shoulder during tough times. Understanding each other's emotions is natural for humans, however, that's an arduous task for machines. This is essentially important when machines are often used for anticipating emotions nowadays and have become a crucial problem for machines for effective human-machine interaction. \par

Emotions can be detected in various ways such as through facial features, behavior, body gestures, physiological signals, and speech. In this work, we focus on Speech emotion recognition (SER), the method of understanding emotions in human speech in particular,
 which has %recently 
received attention due to its potential applications in a broad range of diverse areas, including psychology, healthcare, etc. Various methods have been applied for SER, such as Hidden Markov Model (HMM) \cite{vlasenko2007tuning}, classical machine learning \cite{iliou2009comparison}, deep neural network-based approaches \cite{issa2020speech}. Recent works on SER have also used embeddings from various speech (wav2vec, wav2vec 2.0) and audio (YAMNet, VGGish) pre-trained models (PTMs) as input features for downstream models \cite{keesing2021acoustic}. PTMs can be of varied types, for example, CNN-based (YAMNet, VGGish) and transformer-based (wav2vec, wav2vec 2.0) architectures. They are trained on enormous amounts of data, either in a supervised or self-supervised manner. Self-supervised Learning (SSL) speech PTM embeddings have proven to be state-of-the-art (SOTA) in comparison to other PTM embeddings \cite{wu2022ability} across various speech-related tasks such as SER, speaker count estimation, etc. The wide availability of embeddings from PTMs has made a significant contribution to the progress of SER. \par

However, much focus on speaker recognition PTM embeddings for SER hasn't been given in contrast to speech SSL PTM embeddings such as wav2vec2.0 \cite{pepino2021emotion}, Unispeech-SAT \cite{atmaja2022sentiment}. Pappagari et al. \cite{pappagari2020x} showed the %relationship 
association between speaker and emotion recognition; focusing on the usefulness of speaker recognition PTM embeddings for SER. Nevertheless, for understanding the effectiveness of speaker recognition PTM embeddings for SER, a comparison with other PTM embeddings is required that has been missing in the literature. So, we tackle this research gap by performing a comparative analysis of various PTM embeddings. To summarize, the main contributions of our work are as follows: 
\par

\begin{itemize}

  \item  Comparative analysis of different PTM embeddings (x-vector, ECAPA, wav2vec 2.0, wavLM, Unispeech-SAT) on CREMA-D benchmark. The top performance is achieved by x-vector embeddings among all the other embeddings.
  
  \item Our hypothesis for the performance of x-vector embeddings is that the model learned numerous components of speech such as tone, pitch, %accent, 
  and so on as a result of speaker recognition training. Further refinement of these embeddings with Convolutional Neural Network (CNN) resulted in contributing towards the same, capturing more important representations.
  
  \item Our approach which uses x-vector embeddings and mel-frequency cepstral coefficients (MFCC) as input features to a downstream CNN model achieves comparable performance in terms of accuracy with SOTA works on CREMA-D while being the most lightweight method in terms of the number of parameters involved. See Section \ref{sotaee}.
  
\end{itemize} 

\noindent This paper is divided into five sections. Section \ref{ser} walks through past literature on approaches carried out for SER followed by Section \ref{method} which discusses the PTM embeddings considered for our analysis. In Section \ref{exp}, we provide brief information on the speech emotion corpus considered for our analysis, downstream classifier, training details, and experimental results. Lastly, Section \ref{conc} summarizes the work presented. % and provides potential directions for future work. 

\begin{comment}
\begin{figure}[hbt!]    
\centering
      \includegraphics[width=0.5\textwidth, height=0.4\textwidth]{reference_model_architecture.png}
      \caption{Reference Architecture}
        \label{fig:refnet}     
\end{figure}

\begin{figure}[hbt!]    
\centering
      \includegraphics[width=0.5\textwidth, height=0.4\textwidth]{reference_model_architecture1.png}
      \caption{Reference Architecture1}
        \label{fig:refnet}     
\end{figure} 
\end{comment}

\section{Related work}

\label{ser}

Studies have shed light on SER from the beginning of the 21st century \cite{cowie2001emotion,nogueiras2001speech}. Initial research involved the usage of Hidden Markov Models (HMMs) \cite{schuller2003hidden} and followed by the usage of classical machine learning algorithms with handcrafted features \cite{lee2005toward}. CNN has been applied to SER \cite{huang2014speech} after AlexNet gained popularity in the ImageNet challenge. Huang et al. \cite{huang2014speech} work comprised two phases with CNN: unsupervised learning followed by semi-supervised learning. HMMs were again brought to the limelight by Mao et al. \cite{mao2019revisiting} with certain modifications. They experimented with three different HMM-based architectures, namely, Gaussian Mixture-based HMMs, Subspace Gaussian Mixture-based HMMs, and deep learning-based HMMs. Various CNN architectures such as AlexNet, ResNet, Xception, and different variants of VGG and DenseNet pre-trained on ImageNet for image recognition tasks were leveraged for SER \cite{ottl2020group}. Zhang et al. \cite{zhang2021pre} used AlexNet in conjunction with Bidirectional LSTM and attention mechanism. Fusion of MFCC and features retrieved from pre-trained CNN as input features to LSTM model was put to use by Arano et al. \cite{arano2021old}. Although CNNs were considered to be most suitable for SER, with time, transformers have also earned a spot \cite{lian2021ctnet, lei2022bat}. A model architecture with multiple transformer layers stacked on top of one other was proposed by Wang et al. \cite{wang2021novel}. Heracleous et al. \cite{heracleous2022applying} was the first to use ViT for the purpose of emotion recognition. \par

In recent times, the usage of embeddings from various PTMs like YAMNet, wav2vec, etc., as input features to classifiers for SER \cite{keesing2021acoustic} can be seen. This approach of using embeddings extracted from PTMs has become a widely used method due to the various benefits associated with it, such as saving time and cost involved with training a model from scratch and also boosts in performance as these models are trained on diverse large-scale data and learn nuanced representations of the input data. Speech SSL PTM embeddings have shown superior performance in comparison to embeddings from other PTMs such as YAMnet, VGGish, etc for SER \cite{keesing2021acoustic}. Atmaja et al. \cite{atmaja2022evaluating} gave a comparison of various SOTA speech SSL PTM embeddings by training and evaluating a fully connected network on top of the extracted embeddings. \par

However, research into speaker recognition PTM embeddings hasn't received much attention in comparison to their counterparts for SER, so we work in this direction by comparing different PTM embeddings to evaluate their effectiveness. We hypothesize that PTM initially trained for speaker recognition can be more effective for SER as knowledge gained for speaker recognition such as learning tone, pitch, etc. from speech can be beneficial.

\section{Pre-trained Model Embeddings}
\label{method}

For our analysis, we consider five PTM embeddings: x-vector, ECAPA, wav2vec 2.0, wavLM, and Unispeech-SAT. x-vector \cite{snyder2018x} and its modified version Emphasized Channel Attention, Propagation, and Aggregation (ECAPA) \cite{desplanques2020ecapa} are speaker recognition PTMs. x-vector is a state-of-the-art speaker recognition system and it is a time delay neural network (TDNN) trained end-to-end in a supervised fashion without any handcrafted features for identifying speakers. After the training is completed, it is used for obtaining speaker embeddings from variable-length utterances. x-vector outperforms i-vector, prior speaker recognition system. ECAPA approach made numerous enhancements in the frame and pooling level to the original x-vector model architecture by rectifying its limitations. We use readily available x-vector\footnote{\url{https://huggingface.co/speechbrain/spkrec-xvect-voxceleb}} and ECAPA\footnote{\url{https://huggingface.co/speechbrain/spkrec-ecapa-voxceleb}} models from \textit{HuggingFace}. Combination of voxceleb 1 and voxceleb 2 were used as pre-training data for both the models with the input speech signals sampled as 16KHz single-channel \cite{speechbrain}. We extract embeddings of 512 and 192-dimension from x-vector and ECAPA respectively with the help of \textit{Speechbrain} \cite{speechbrain} library.
\par

We follow Speech processing Universal PERformance Benchmark (SUPERB) \cite{yang21c_interspeech} during consideration of speech SSL PTMs (wav2vec 2.0, wavLM, Unispeech-SAT). SUPERB evaluates features from SSL PTMs across a wide range of tasks, such as speaker identification, SER, speech recognition, voice separation, and so on. WavLM \cite{chen2022wavlm} outperformed all its competitors except Unispeech-SAT. It is proposed as a generalized model for solving various posterior speech-related tasks. During pre-training, wavLM learns both speech prediction and denoising in conjunction which aids in understanding the multi-dimensional information such as speaker identity, content, etc. embedded in speech. In contrast, pre-training for the Unispeech-SAT was done in a speaker-aware way. It is a contrastive loss based multitask learning model. For our experiments, we utilize wavLM base+\footnote{\url{https://huggingface.co/docs/transformers/model_doc/wavlm}} and UniSpeech-SAT base\footnote{\url{https://huggingface.co/docs/transformers/model_doc/unispeech-sat}} version. WavLM base+ has 12 transformer-encoder layers and pre-training was done on 94k hours of data from several speech datasets such as LibriLight, VoxPopuli, and GigaSpeech whereas the Unispeech-SAT base version was pre-trained on 960 hours of Librispeech. \par

The performance of wav2vec 2.0 is not as high in comparison with wavLM and Unispeech-SAT on SUPERB. However, previous researchers have made use of its embeddings for SER \cite{pepino2021emotion} and has proven to be effective so we add it to our experiments. wav2vec 2.0 was pre-trained in a self-supervised manner on Librispeech. We use base\footnote{\url{https://huggingface.co/facebook/wav2vec2-base}} version that contains 12 transformer blocks. The final hidden states are retrieved from wavLM, UniSpeech-SAT, and wav2vec 2.0 and transformed to a vector of 768-dimension for each audio file to be utilized as input features for the downstream classifier using pooling average. Sampling is done at 16KHz for each audio file to be provided as input for the SSL PTMs. 

\begin{comment}
    
\textcolor{red}{x-vector is a state-of-the-art speaker recognition system that has gained widespread popularity in recent years. It is a deep neural network-based approach that extracts speaker embeddings from variable-length utterances, allowing for speaker identification and verification tasks. The x-vector system is trained end-to-end, requiring no hand-crafted feature engineering or model tuning. The x-vector model comprises several layers of neural networks, including convolutional layers, time-delay neural networks, and feed-forward layers. These layers extract high-level speaker characteristics from the input speech signals, which are then mapped to a low-dimensional speaker embedding space. The resulting embeddings represent the speaker's unique identity, allowing for speaker recognition tasks. One of the key advantages of the x-vector system is its ability to generalize well to unseen speakers, making it an ideal solution for large-scale speaker recognition applications. Additionally, x-vector has achieved state-of-the-art performance on various speaker recognition benchmarks, outperforming traditional speaker recognition systems.}

\end{comment}

\section{Experiments}
\label{exp}
\subsection{Benchmark Speech Emotion Corpus}
We select Crowd-sourced emotional multimodal actors dataset (CREMA-D) \cite{cao2014crema} for our experiments as it provides a rich data source for SER due to the variations in ages and ethnicities across speakers. It acts as high-quality benchmark data for training and evaluating machine learning models. It is a gender-balanced database in English. It contains 7442 utterances from 91 different speakers across six emotions: Anger, Happiness, Sadness, Fear, Disgust, and Neutral. The audio clips include 48 male and 43 female artists. They spoke from a list of twelve sentences.

\subsection{Downstream Model}

\begin{figure}[hbt!]    
\centering
      \includegraphics[width=0.45\textwidth, height=0.35\textwidth]{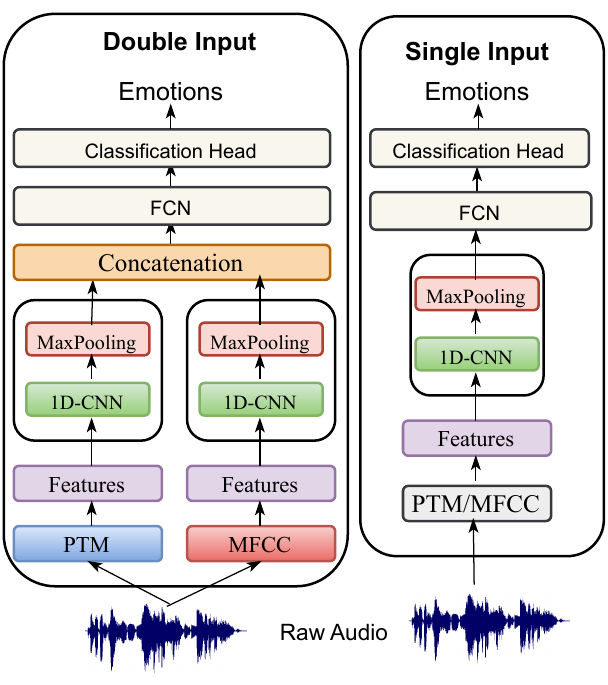}
      \caption{Proposed Model architecture for Double Input i.e embeddings from PTM + MFCC as input features; Single input i.e embeddings from PTM/MFCC as input features}
        \label{fig:archi}     
\end{figure}

For the first set of experiments, which involves a comparison of PTM embeddings, we use 1D-CNN on top of the embeddings from PTMs followed by a Fully Connected Network (FCN) as shown in Figure \ref{fig:archi}. \textit{Softmax} function is used in the output layer that outputs the probabilities for different emotions. The same modeling approach is followed across all the PTM embeddings. Secondly, for double input (PTM + MFCC features), a similar modeling technique is used with 1D-CNN applied on both inputs and then concatenated together. For extracting MFCC from raw audio waveforms, we use \textit{Librosa}\footnote{\url{https://librosa.org/doc/main/generated/librosa.feature.mfcc.html}} library. The modeling approach remains same for different variations of input. We utilize \textit{Tensorflow} library for our experiments.\par

Models are trained in a 5-fold fashion, with four folds utilized for training and one fold maintained for testing. Information regarding the hyperparameters set can be found in Table \ref{tab:hyper}. The hyperparameters other than those mentioned are kept as default. We also use early stopping and learning rate decay during training.

\begin{table}[hbt!]
\centering
\caption{Hyperparameter Details}
\label{tab:hyper}
%\addtolength{\tabcolsep}{-5pt}
\begin{tabular}{ll}   %{|l|l|}
\hline \hline
Number of Kernels for 1D-CNN & 32  \\  
Kernel Size for 1D-CNN & 3\\
Number of Neurons for each layer of FCN & 200, 90, 56\\
Activation Function in Intermediate Layers & ReLU\\
Training epochs & 50 \\
Optimizer & Rectified Adam \\
Learning  Rate & 1e-3\\
Batch Size & 32\\
\hline \hline
\end{tabular}
\end{table}

\subsection{Experimental Results}

\begin{figure}[t!]
    \centering
    \subfloat[x-vector]{{\includegraphics[width=0.52\linewidth, height =3.2cm ]{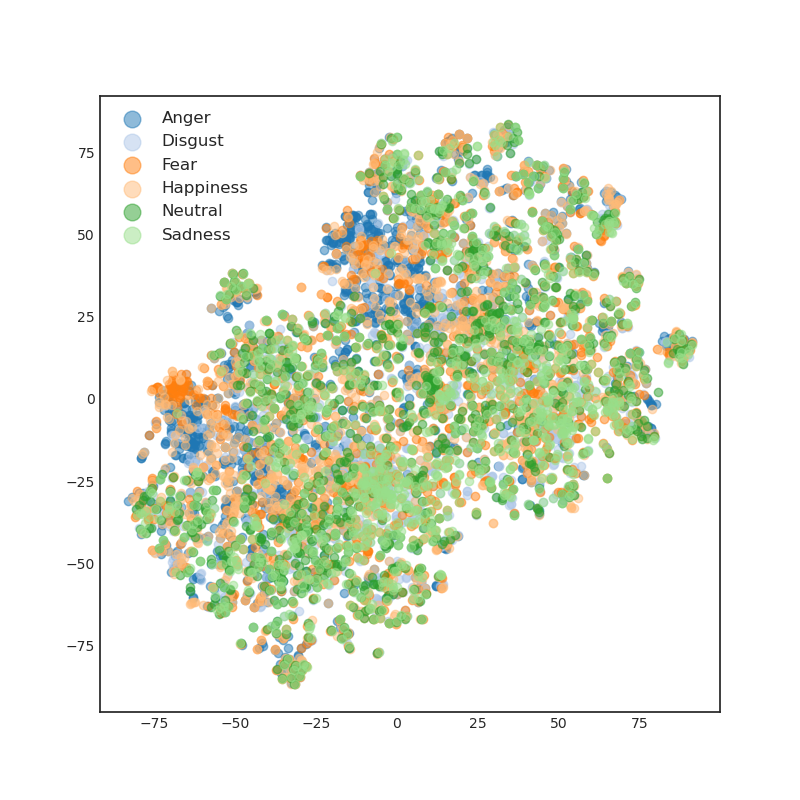}}\label{fig:xvec}}
    \subfloat[ECAPA]
    {{\includegraphics[width=0.52\linewidth,height =3.2cm ]{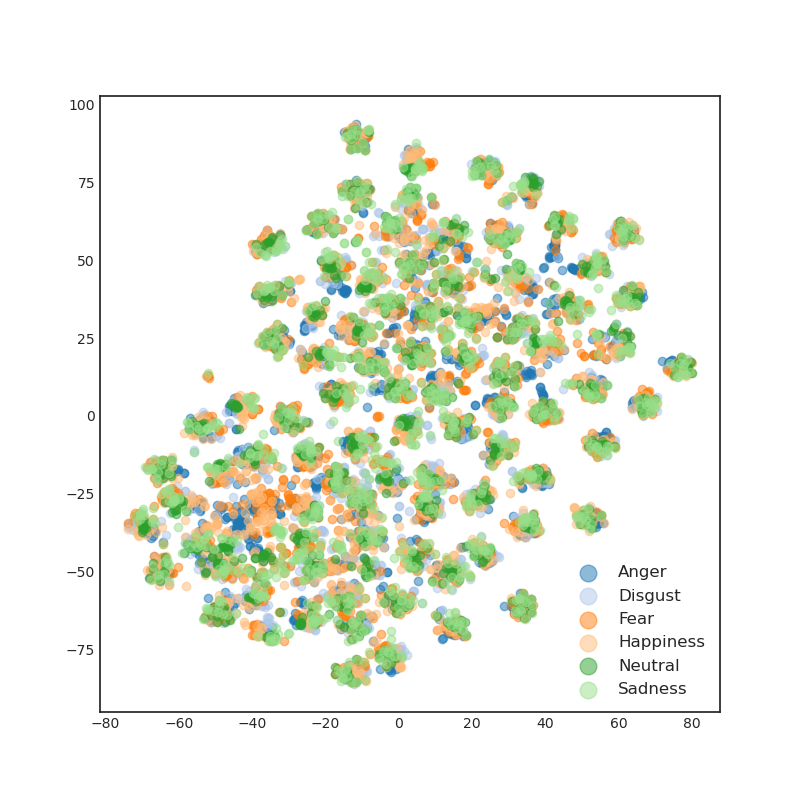}}\label{fig:ecapa}}\\
    \subfloat[wav2vec 2.0]
    {{\includegraphics[width=0.52\linewidth,height =3.2cm ]{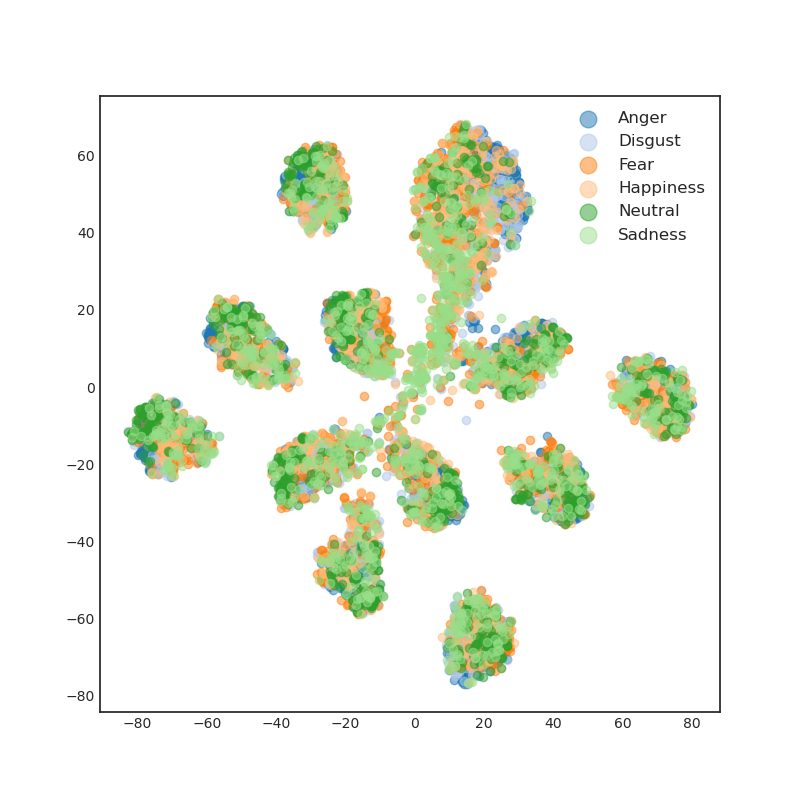}}\label{fig:wav2vec2}}
    \subfloat[wavLM]
    {{\includegraphics[width=0.52\linewidth,height=3.2cm ]{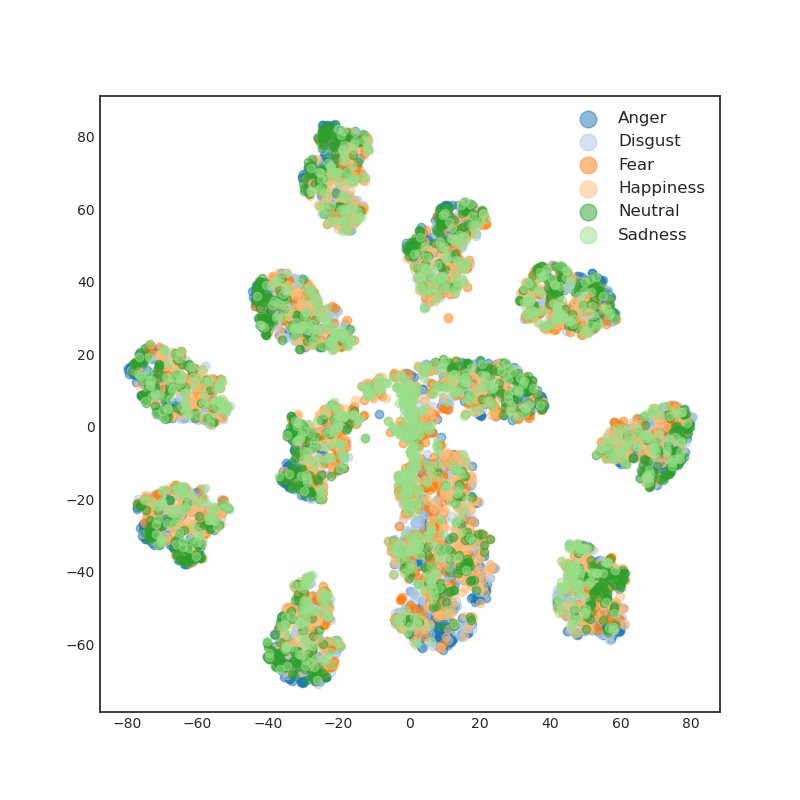}}\label{fig:wavlm}}\\
    \subfloat[Unispeech-SAT]
    {{\includegraphics[width=0.52\linewidth,height =3.2cm ]{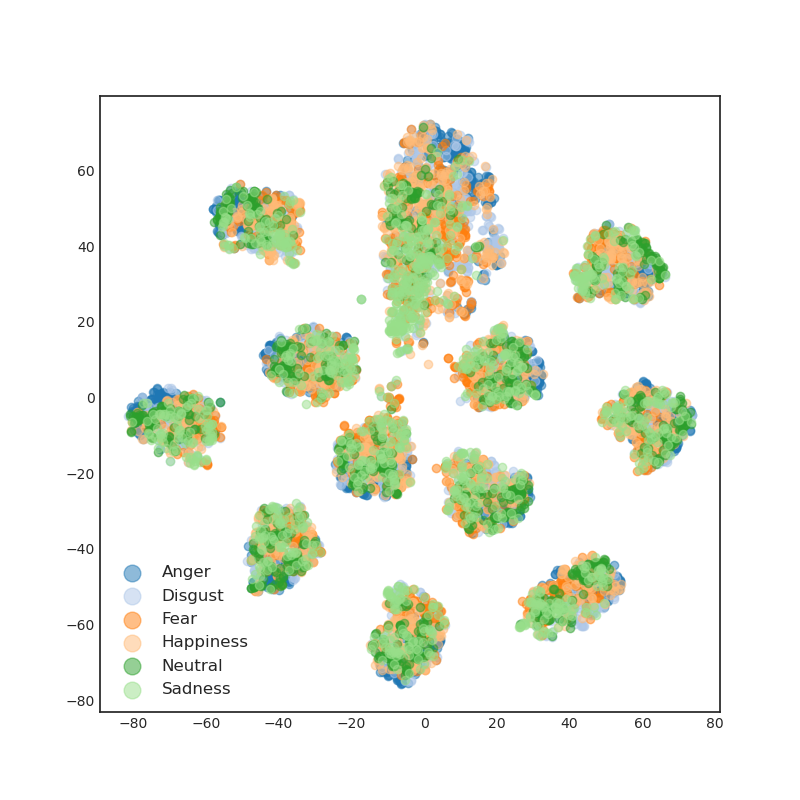}}\label{fig:uni}}
    \subfloat[x-vector\_TESS]{{\includegraphics[width=0.52\linewidth, height =3.2cm ]{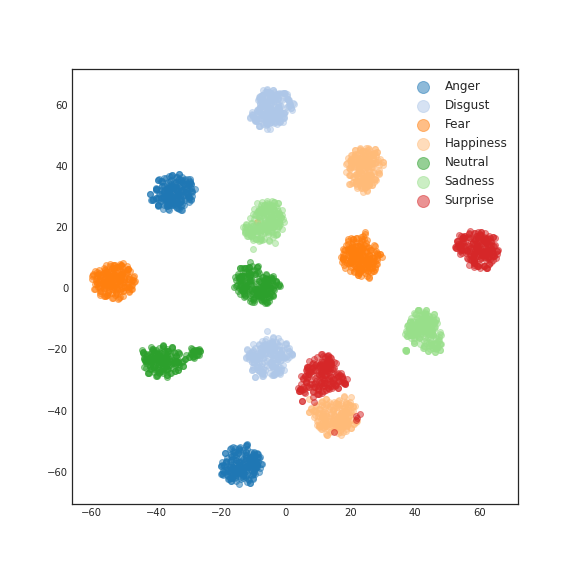}}\label{fig:xvectess}}
   \caption{Figure \ref{fig:xvec}, \ref{fig:ecapa}, \ref{fig:wav2vec2}, \ref{fig:wavlm}, \ref{fig:uni} represents the t-SNE plots of PTM embeddings on CREMA-D and Figure \ref{fig:xvectess} represents the t-SNE plot of x-vector embeddings on TESS}
\label{fig:tsne}
\end{figure}

We compare the performance of PTM embeddings by training and evaluating them with the downstream model. The results are shown in Table \ref{tab:embed}. The performance of baseline MFCC features and wav2vec 2.0 embeddings is not significantly different which points towards wav2vec 2.0 embeddings being unable to capture important information in comparison to its other PTM counterparts. Unispeech-SAT achieves higher performance than ECAPA and other speech SSL PTM embeddings (wav2vec 2.0, wavLM). This can be resultant of its speaker-aware pre-training. However, ECAPA performs better than the other two speech SSL PTM embeddings. Out of all the PTM embeddings, the model trained on x-vector embeddings performed the best with an accuracy of 68.19\%. This validates our hypothesis that models pre-trained for speaker recognition learns about tone, pitch, and other characteristics which are useful for SER. We also plot t-SNE plots of the raw embeddings retrieved from the PTMs and is shown in Figure \ref{fig:tsne}. These figures follow up the results presented in Table \ref{tab:embed}, with slight cluster formation in accordance to different emotions that can be seen for x-vector embeddings in comparison to others. \par

We extended our experiments by combining PTM embeddings with MFCC as input features, and the results are shown in Table \ref{tab:embmf}. Except for Unispeech-SAT, combining MFCC with PTM embeddings results in increased accuracy. 

\subsection{Comparison to State-of-the-art}
\label{sotaee}

\textbf{Accuracy:} We compare our top-performing model (x-vector + MFCC) against SOTA works presented in Table \ref{tab:sota}. As seen, our approach is able to attain comparable performance against SOTA methods. \par

\noindent \textbf{Lightweight:} We compare our model in terms of the number of parameters with representative SOTA works which have mentioned the parameters in their models. We have not compared our model to CLAP \cite{dhamyal2022describing} as it involves additional modality i.e. descriptions of the speech emotions and we are only comparing with works that have worked only with speech as input. ViT \cite{gong21b_interspeech} has 75.7M parameters while SepTr \cite{ristea22_interspeech} has 9.4M parameters. SepTr + LeRaC \cite{croitoru2022lerac} also contains a similar number of parameters as SepTr. Our model has 1.79M parameters which are far lesser parameters than the models mentioned above. During the training phase, it requires only 1 second per epoch on P100 GPU. For fusion\_cat\_xwc \cite{wu2022ability}, it's difficult to quantify the exact number of parameters involved as it is a work carried out for HEAR \cite{turian2022hear}. It consists of an ensemble of several speech SSL PTMs with no fine-tuning followed by a downstream Multilayer Perceptron classifier.

%Total Params: 1,791,184
%Trainable params: 1,790,784
%Non-trainable params: 400

\begin{table}[hbt!]
\centering
\caption{Performance of different PTM embeddings; MFCC are taken as baseline input features and all the scores are average of 5-folds}
\label{tab:embed}
%\addtolength{\tabcolsep}{-5pt}
\begin{tabular}{ll}   %{|l|l|}
\hline\hline
\textbf{Input Features} & \textbf{Accuracy (\%)}\\\hline
MFCC & 47.44\\
x-vector &  \textbf{68.19}\\ 
ECAPA & 58.91\\ 
wav2vec 2.0 & 48.12 \\ 
wavLM &  55.68\\ 
Unispeech-SAT & 67.27\\  
\hline\hline
\end{tabular}
\end{table}

\begin{table}[hbt!]
\centering
\caption{Performance achieved by combination of PTM embeddings + MFCC as Input Features; all the scores are average of 5-folds}
\label{tab:embmf}
%\addtolength{\tabcolsep}{-5pt}
\begin{tabular}{ll}   %{|l|l|}
\hline\hline
\textbf{Input Features} & \textbf{Accuracy (\%)}\\\hline
x-vector + MFCC & \textbf{70.53} \\ 
ECAPA + MFCC & 63.02 \\ 
wav2vec 2.0 + MFCC &  58.13\\ 
wavLM + MFCC & 57.52\\ 
Unispeech-SAT + MFCC & 67.01\\ 
\hline\hline
\end{tabular}
\end{table}

\begin{table}[hbt!]
\centering
\caption{Comparison of proposed approaches (CNN(x-vector), CNN(x-vector, MFCC)) with SOTA works on CREMA-D benchmark; CNN(x-vector), CNN(x-vector, MFCC) represents CNN model with x-vector embeddings and x-vector embeddings + MFCC as input features respectively}
\label{tab:sota}
%\addtolength{\tabcolsep}{-5pt}
\begin{tabular}{lll}   %{|l|l|}
\hline\hline
\textbf{Method} & \textbf{Accuracy (\%)} & \textbf{Year}\\\hline
GRU \cite{shukla2020visually} & 55.01 &  2020\\ 
GAN \cite{he2020image2audio} & 58.71 &  2020\\ 
ResNet-18 \cite{georgescu2020non} & 65.15 &  2020 \\ 
ViT \cite{gong21b_interspeech} & 67.81 & 2020 \\ 
ResNet-18 ensemble \cite{ristea2021self} & 68.12 & 2021\\ 
CNN(x-vector) & 68.19 & \\
SepTr-VH \cite{ristea22_interspeech} & 70.47 & 2022\\ 
CNN(x-vector, MFCC) & 70.53 & \\ 
SepTr + LeRaC \cite{croitoru2022lerac} & 70.95 & 2022\\
CLAP \cite{dhamyal2022describing} & 72.56 & 2022\\
fusion\_cat\_xwc \cite{wu2022ability} & \textbf{74.70} & 2022\\
\hline\hline
\end{tabular}
\end{table}

\begin{table}[hbt!]
\centering
\caption{Performance of proposed approach (CNN(x-vector, MFCC))  and its comparison with previous methods on TESS}
\label{tab:tess}
%\addtolength{\tabcolsep}{-5pt}
\begin{tabular}{llll}   %{|l|l|}
\hline\hline
\textbf{Approach} & \textbf{Accuracy (\%)} & \textbf{Year} \\\hline 
CNN \cite{huang2019human}& 85 & 2019\\ 
CNN \cite{choudhary2022speech} & 97.1 & 2022\\ 
CNN(x-vector, MFCC) & 99.82 & \\ 
IMEMD-CRNN \cite{sun2022speech} & \textbf{100} & 2022\\ 
\hline\hline
\end{tabular}
\end{table}

\subsection{Additional Experiment}

Furthermore, we evaluated our model (CNN(x-vector, MFCC)) on an additional speech emotion corpus i.e. Toronto Emotional Speech Set (TESS) \cite{utorontoTorontoEmotional} to evaluate the effectiveness of our model. TESS is an English database consisting of 2800 utterances with 2 female speakers spreading across seven emotions: Anger, Happiness, Sadness, Fear, Disgust, Neutral, and Surprise. We use the same set of hyperparameters as given in Table \ref{tab:hyper}. We follow a 5-fold procedure for training our model and report the average accuracy score for 5-folds in comparison with previous works which is presented in Table \ref{tab:tess}. t-SNE plot of raw embeddings extracted from x-vector is shown in Figure \ref{fig:xvectess}, with clear segregation of clusters is observed based on different emotions. These results show the effectiveness of our model for SER, especially, embeddings from x-vector as input features. 

\section{Conclusions}
\label{conc}

%With the presence of a wide range of PTMs SER has made sufficient progress.PTMs can be of varied types. Past literature has used embeddings from different PTMs as input features to lower-level models for SER. Speech SSL PTM embeddings are the most preferred option due to their SOTA performance across various speech-related tasks including SER. Using embeddings from PTMs is a recent trend for SER. However, previous work has paid less focus to the usage of speaker recognition embeddings for SER. So in this work, we rectify this research gap and for evaluating their usefulness,
In this work, we perform a comprehensive comparative study of five  PTMs (x-vector, ECAPA, wav2vec 2.0, wavLM, Unispeech-SAT) for analyzing their performance for SER. x-vector embeddings fared the best in comparison to the other PTM embeddings, which might be due to its training for speaker recognition, which leads to the capture of many important components of speech such as tone, pitch, and so on. x-vector embeddings and MFCC as input features based downstream CNN approach attained comparable performance in accordance with accuracy and also the most lightweight technique on CREMA-D. The results of this study can aid in selecting appropriate embeddings for tasks related to SER and serve as a baseline approach for future research in this direction.

\bibliographystyle{IEEEtran}
\bibliography{mybib}

\begin{thebibliography}{10}
\providecommand{\url}[1]{#1}
\csname url@samestyle\endcsname
\providecommand{\newblock}{\relax}
\providecommand{\bibinfo}[2]{#2}
\providecommand{\BIBentrySTDinterwordspacing}{\spaceskip=0pt\relax}
\providecommand{\BIBentryALTinterwordstretchfactor}{4}
\providecommand{\BIBentryALTinterwordspacing}{\spaceskip=\fontdimen2\font plus
\BIBentryALTinterwordstretchfactor\fontdimen3\font minus
  \fontdimen4\font\relax}
\providecommand{\BIBforeignlanguage}[2]{{%
\expandafter\ifx\csname l@#1\endcsname\relax
\typeout{** WARNING: IEEEtran.bst: No hyphenation pattern has been}%
\typeout{** loaded for the language `#1'. Using the pattern for}%
\typeout{** the default language instead.}%
\else
\language=\csname l@#1\endcsname
\fi
#2}}
\providecommand{\BIBdecl}{\relax}
\BIBdecl

\bibitem{vlasenko2007tuning}
B.~Vlasenko and A.~Wendemuth, ``Tuning hidden markov model for speech emotion
  recognition,'' \emph{Fortschritte der akustik}, vol.~33, no.~1, p. 317, 2007.

\bibitem{iliou2009comparison}
T.~Iliou and C.-N. Anagnostopoulos, ``Comparison of different classifiers for
  emotion recognition,'' in \emph{2009 13th Panhellenic Conference on
  Informatics}.\hskip 1em plus 0.5em minus 0.4em\relax IEEE, 2009, pp.
  102--106.

\bibitem{issa2020speech}
D.~Issa, M.~F. Demirci, and A.~Yazici, ``Speech emotion recognition with deep
  convolutional neural networks,'' \emph{Biomedical Signal Processing and
  Control}, vol.~59, p. 101894, 2020.

\bibitem{keesing2021acoustic}
A.~Keesing, Y.~S. Koh, and M.~Witbrock, ``Acoustic features and neural
  representations for categorical emotion recognition from speech.'' in
  \emph{Interspeech}, 2021, pp. 3415--3419.

\bibitem{wu2022ability}
T.-Y. Wu, C.-A. Li, T.-H. Lin, T.-Y. Hsu, and H.-Y. Lee, ``The ability of
  self-supervised speech models for audio representations,'' \emph{arXiv
  preprint arXiv:2209.12900}, 2022.

\bibitem{pepino2021emotion}
L.~Pepino, P.~Riera, and L.~Ferrer, ``Emotion recognition from speech using
  wav2vec 2.0 embeddings,'' \emph{Proc. Interspeech 2021}, pp. 3400--3404,
  2021.

\bibitem{atmaja2022sentiment}
B.~T. Atmaja and A.~Sasou, ``Sentiment analysis and emotion recognition from
  speech using universal speech representations,'' \emph{Sensors}, vol.~22,
  no.~17, p. 6369, 2022.

\bibitem{pappagari2020x}
R.~Pappagari, T.~Wang, J.~Villalba, N.~Chen, and N.~Dehak, ``x-vectors meet
  emotions: A study on dependencies between emotion and speaker recognition,''
  in \emph{ICASSP 2020-2020 IEEE International Conference on Acoustics, Speech
  and Signal Processing (ICASSP)}.\hskip 1em plus 0.5em minus 0.4em\relax IEEE,
  2020, pp. 7169--7173.

\bibitem{cowie2001emotion}
R.~Cowie, E.~Douglas-Cowie, N.~Tsapatsoulis, G.~Votsis, S.~Kollias, W.~Fellenz,
  and J.~G. Taylor, ``Emotion recognition in human-computer interaction,''
  \emph{IEEE Signal processing magazine}, vol.~18, no.~1, pp. 32--80, 2001.

\bibitem{nogueiras2001speech}
A.~Nogueiras, A.~Moreno, A.~Bonafonte, and J.~B. Mari{\~n}o, ``Speech emotion
  recognition using hidden markov models,'' in \emph{Seventh European
  conference on speech communication and technology}, 2001.

\bibitem{schuller2003hidden}
B.~Schuller, G.~Rigoll, and M.~Lang, ``Hidden markov model-based speech emotion
  recognition,'' in \emph{2003 IEEE International Conference on Acoustics,
  Speech, and Signal Processing, 2003. Proceedings.(ICASSP'03).}, vol.~2.\hskip
  1em plus 0.5em minus 0.4em\relax Ieee, 2003, pp. II--1.

\bibitem{lee2005toward}
C.~M. Lee and S.~S. Narayanan, ``Toward detecting emotions in spoken dialogs,''
  \emph{IEEE transactions on speech and audio processing}, vol.~13, no.~2, pp.
  293--303, 2005.

\bibitem{huang2014speech}
Z.~Huang, M.~Dong, Q.~Mao, and Y.~Zhan, ``Speech emotion recognition using
  cnn,'' in \emph{Proceedings of the 22nd ACM international conference on
  Multimedia}, 2014, pp. 801--804.

\bibitem{mao2019revisiting}
S.~Mao, D.~Tao, G.~Zhang, P.~Ching, and T.~Lee, ``Revisiting hidden markov
  models for speech emotion recognition,'' in \emph{ICASSP 2019-2019 IEEE
  International Conference on Acoustics, Speech and Signal Processing
  (ICASSP)}.\hskip 1em plus 0.5em minus 0.4em\relax IEEE, 2019, pp. 6715--6719.

\bibitem{ottl2020group}
S.~Ottl, S.~Amiriparian, M.~Gerczuk, V.~Karas, and B.~Schuller, ``Group-level
  speech emotion recognition utilising deep spectrum features,'' in
  \emph{Proceedings of the 2020 International Conference on Multimodal
  Interaction}, 2020, pp. 821--826.

\bibitem{zhang2021pre}
H.~Zhang, R.~Gou, J.~Shang, F.~Shen, Y.~Wu, and G.~Dai, ``Pre-trained deep
  convolution neural network model with attention for speech emotion
  recognition,'' \emph{Frontiers in Physiology}, vol.~12, p. 643202, 2021.

\bibitem{arano2021old}
K.~A. Ara{\~n}o, P.~Gloor, C.~Orsenigo, and C.~Vercellis, ``When old meets new:
  emotion recognition from speech signals,'' \emph{Cognitive Computation},
  vol.~13, pp. 771--783, 2021.

\bibitem{lian2021ctnet}
Z.~Lian, B.~Liu, and J.~Tao, ``Ctnet: Conversational transformer network for
  emotion recognition,'' \emph{IEEE/ACM Transactions on Audio, Speech, and
  Language Processing}, vol.~29, pp. 985--1000, 2021.

\bibitem{lei2022bat}
J.~Lei, X.~Zhu, and Y.~Wang, ``Bat: Block and token self-attention for speech
  emotion recognition,'' \emph{Neural Networks}, vol. 156, pp. 67--80, 2022.

\bibitem{wang2021novel}
X.~Wang, M.~Wang, W.~Qi, W.~Su, X.~Wang, and H.~Zhou, ``A novel end-to-end
  speech emotion recognition network with stacked transformer layers,'' in
  \emph{ICASSP 2021-2021 IEEE International Conference on Acoustics, Speech and
  Signal Processing (ICASSP)}.\hskip 1em plus 0.5em minus 0.4em\relax IEEE,
  2021, pp. 6289--6293.

\bibitem{heracleous2022applying}
P.~Heracleous, S.~Fukayama, J.~Ogata, and Y.~Mohammad, ``Applying generative
  adversarial networks and vision transformers in speech emotion recognition,''
  in \emph{HCI International 2022-Late Breaking Papers. Multimodality in
  Advanced Interaction Environments: 24th International Conference on
  Human-Computer Interaction, HCII 2022, Virtual Event, June 26--July 1, 2022,
  Proceedings}.\hskip 1em plus 0.5em minus 0.4em\relax Springer, 2022, pp.
  67--75.

\bibitem{atmaja2022evaluating}
B.~T. Atmaja and A.~Sasou, ``Evaluating self-supervised speech representations
  for speech emotion recognition,'' \emph{IEEE Access}, vol.~10, pp.
  124\,396--124\,407, 2022.

\bibitem{snyder2018x}
D.~Snyder, D.~Garcia-Romero, G.~Sell, D.~Povey, and S.~Khudanpur, ``X-vectors:
  Robust dnn embeddings for speaker recognition,'' in \emph{2018 IEEE
  international conference on acoustics, speech and signal processing
  (ICASSP)}.\hskip 1em plus 0.5em minus 0.4em\relax IEEE, 2018, pp. 5329--5333.

\bibitem{desplanques2020ecapa}
B.~Desplanques, J.~Thienpondt, and K.~Demuynck, ``Ecapa-tdnn: Emphasized
  channel attention, propagation and aggregation in tdnn based speaker
  verification,'' 2020.

\bibitem{speechbrain}
M.~Ravanelli, T.~Parcollet, P.~Plantinga, A.~Rouhe, S.~Cornell, L.~Lugosch,
  C.~Subakan, N.~Dawalatabad, A.~Heba, J.~Zhong, J.-C. Chou, S.-L. Yeh, S.-W.
  Fu, C.-F. Liao, E.~Rastorgueva, F.~Grondin, W.~Aris, H.~Na, Y.~Gao, R.~D.
  Mori, and Y.~Bengio, ``{SpeechBrain}: A general-purpose speech toolkit,''
  2021, arXiv:2106.04624.

\bibitem{yang21c_interspeech}
S.~wen Yang, P.-H. Chi, Y.-S. Chuang, C.-I.~J. Lai, K.~Lakhotia, Y.~Y. Lin,
  A.~T. Liu, J.~Shi, X.~Chang, G.-T. Lin, T.-H. Huang, W.-C. Tseng, K.~tik Lee,
  D.-R. Liu, Z.~Huang, S.~Dong, S.-W. Li, S.~Watanabe, A.~Mohamed, and
  H.~yi~Lee, ``{SUPERB: Speech Processing Universal PERformance Benchmark},''
  in \emph{Proc. Interspeech 2021}, 2021, pp. 1194--1198.

\bibitem{chen2022wavlm}
S.~Chen, C.~Wang, Z.~Chen, Y.~Wu, S.~Liu, Z.~Chen, J.~Li, N.~Kanda,
  T.~Yoshioka, X.~Xiao \emph{et~al.}, ``Wavlm: Large-scale self-supervised
  pre-training for full stack speech processing,'' \emph{IEEE Journal of
  Selected Topics in Signal Processing}, vol.~16, no.~6, pp. 1505--1518, 2022.

\bibitem{cao2014crema}
H.~Cao, D.~G. Cooper, M.~K. Keutmann, R.~C. Gur, A.~Nenkova, and R.~Verma,
  ``Crema-d: Crowd-sourced emotional multimodal actors dataset,'' \emph{IEEE
  transactions on affective computing}, vol.~5, no.~4, pp. 377--390, 2014.

\bibitem{dhamyal2022describing}
H.~Dhamyal, B.~Elizalde, S.~Deshmukh, H.~Wang, B.~Raj, and R.~Singh,
  ``Describing emotions with acoustic property prompts for speech emotion
  recognition,'' \emph{arXiv preprint arXiv:2211.07737}, 2022.

\bibitem{gong21b_interspeech}
Y.~Gong, Y.-A. Chung, and J.~Glass, ``{AST: Audio Spectrogram Transformer},''
  in \emph{Proc. Interspeech 2021}, 2021, pp. 571--575.

\bibitem{ristea22_interspeech}
N.~C. Ristea, R.~T. Ionescu, and F.~S. Khan, ``{SepTr: Separable Transformer
  for Audio Spectrogram Processing},'' in \emph{Proc. Interspeech 2022}, 2022,
  pp. 4103--4107.

\bibitem{croitoru2022lerac}
F.-A. Croitoru, N.-C. Ristea, R.~T. Ionescu, and N.~Sebe, ``Lerac: Learning
  rate curriculum,'' \emph{arXiv preprint arXiv:2205.09180}, 2022.

\bibitem{turian2022hear}
J.~Turian, J.~Shier, H.~R. Khan, B.~Raj, B.~W. Schuller, C.~J. Steinmetz,
  C.~Malloy, G.~Tzanetakis, G.~Velarde, K.~McNally \emph{et~al.}, ``Hear:
  Holistic evaluation of audio representations,'' in \emph{NeurIPS 2021
  Competitions and Demonstrations Track}.\hskip 1em plus 0.5em minus
  0.4em\relax PMLR, 2022, pp. 125--145.

\bibitem{shukla2020visually}
A.~Shukla, K.~Vougioukas, P.~Ma, S.~Petridis, and M.~Pantic, ``Visually guided
  self supervised learning of speech representations,'' in \emph{ICASSP
  2020-2020 IEEE International Conference on Acoustics, Speech and Signal
  Processing (ICASSP)}.\hskip 1em plus 0.5em minus 0.4em\relax IEEE, 2020, pp.
  6299--6303.

\bibitem{he2020image2audio}
G.~He, X.~Liu, F.~Fan, and J.~You, ``Image2audio: Facilitating semi-supervised
  audio emotion recognition with facial expression image,'' in
  \emph{Proceedings of the IEEE/CVF Conference on Computer Vision and Pattern
  Recognition Workshops}, 2020, pp. 912--913.

\bibitem{georgescu2020non}
M.-I. Georgescu, R.~T. Ionescu, N.-C. Ristea, and N.~Sebe, ``Non-linear neurons
  with human-like apical dendrite activations,'' \emph{arXiv preprint
  arXiv:2003.03229}, 2020.

\bibitem{ristea2021self}
N.-C. Ristea and R.~T. Ionescu, ``Self-paced ensemble learning for speech and
  audio classification,'' \emph{arXiv preprint arXiv:2103.11988}, 2021.

\bibitem{huang2019human}
A.~Huang and P.~Bao, ``Human vocal sentiment analysis,'' \emph{arXiv preprint
  arXiv:1905.08632}, 2019.

\bibitem{choudhary2022speech}
R.~R. Choudhary, G.~Meena, and K.~K. Mohbey, ``Speech emotion based sentiment
  recognition using deep neural networks,'' in \emph{Journal of Physics:
  Conference Series}, vol. 2236, no.~1.\hskip 1em plus 0.5em minus 0.4em\relax
  IOP Publishing, 2022, p. 012003.

\bibitem{sun2022speech}
C.~Sun, H.~Li, and L.~Ma, ``Speech emotion recognition based on improved
  masking emd and convolutional recurrent neural network,'' \emph{Frontiers in
  Psychology}, vol.~13, 2022.

\bibitem{utorontoTorontoEmotional}
M.~K. P.-F. Kate~Dupuis, ``{T}oronto emotional speech set ({T}{E}{S}{S}) |
  {T}{S}pace {R}epository --- tspace.library.utoronto.ca,''
  \url{https://tspace.library.utoronto.ca/handle/1807/24487}, 2010, [Accessed
  06-Nov-2022].

\end{thebibliography}

\end{document}